\documentclass[graybox]{svmult}

\usepackage{url}
\usepackage{hyperref}
\usepackage{mathptmx}       
\usepackage{helvet}         
\usepackage{courier}        
\usepackage{type1cm}        
\usepackage{wrapfig}                            
\usepackage{makeidx}         
\usepackage{graphicx}        
\usepackage{amsmath}                             
\usepackage{multicol}        
\usepackage[bottom]{footmisc}

\usepackage{array}

\usepackage{float}
\usepackage{graphicx,caption}
\usepackage{geometry}
\makeindex             

\title*{Effect of final state interactions on neutrino energy reconstruction at DUNE}

\author{Sabeeha Naaz, Anupam Yadav, Jyotsna Singh and R. B. Singh}
\institute{Sabeeha Naaz, Anupam Yadav and Dr. Jyotsna Singh \at University of Lucknow, Department of Physics,Lucknow-226007,India 
\email{sabeehanaaz0786@gmail.com}
\email{yadavanupam975@gmail.com}
\email{singh.jyotsnalu@gmail.com}
\email{rajendrasinghrb@gmail.com}}
\begin{document}
\maketitle
\vspace{-25mm}
\abstract:{We quantitatively study the percentage of fake events present in CCQE and CCRes
interactions and the impact of final state interactions on the neutrino oscillation
parameters at DUNE. The effect of final-state interactions for DUNE
oscillation physics is analysed in an ideal and realistic detector scenario. 
Resonance interaction will be the most dominant interaction in the oscillation sensitive region of DUNE.  The
$\nu _{\mu}$-disappearance oscillation channel is studied using LAr detector. We find that nuclear effects and
detector threshold plays a significant role in CCQE and CCRes interactions and these
nuclear effects induces a significant bias in the determination of atmospheric neutrino oscillation
parameters. The impression of nuclear effects on the determination of $\theta _{23}$ is quantified in this work, which will
help in reducing the systematic error at DUNE.}\\

Keywords: Final State interaction, resonance, oscillation physics, fake events.

\section{Introduction:}
 The neutrino nucleon interaction study is vital for any long baseline neutrino experiments. In these
experiments the neutrino interacts with a nucleon present in the target nucleus of the detector. As
these nucleons are not free hence nuclear effects can not be ignored while studying the physics
related to neutrino nucleon interaction. In neutrino oscillation experiments the neutrino interaction
with nuclear targets of the detector provides a quantitative handle to neutrino oscillation physics.
Materials with high atomic number are used as detector materials in order to increase the
interaction rates. Nuclear effects are highly intertwined in the nuclear targets i.e. nuclear fermi
motion effects, uncertainties from the binding energy, multinuclear correlation and final state
interactions of produced hadrons in different interaction channels.\\
A quantitative understanding with sufficient accuracy of neutrino nucleon interaction cross section
is essential for the extraction of neutrino energy and neutrino oscillation parameters from the event
rate recorded by the detector. The neutrino beam generated in neutrino oscillation experiments is
not monochromatic on the contrary it is sufficiently widespread. According to
neutrino experiment selected i.e. NoVA \cite{1}, T2K \cite{2}, MINOS \cite{3}, DUNE \cite{6} etc., the energy of neutrino beam
varies from few MeV to several tens of GeV. In this energy region the basic interaction modes are
quasi elastic scattering (QE), resonance (Res), deep inelastic scattering (DIS), two body scattering on a
pair of correlated nucleons and coherent pion production. Below 1 GeV quasi elastic scattering is
the dominant mode of interaction but as we increase in energy resonance and DIS processes over
shadows the QE interactions. Superposition of different interaction mechanism in a given neutrino
beam complicates the reconstruction of neutrino energy whose precise knowledge is the central
element of neutrino oscillation physics \cite{4,5,6,7}.\\
The process of neutrino energy reconstruction is very difficult and entirely different from other
leptons. Neutrino energy reconstruction can be performed by two methods: (i) Kinematic method:
uses the kinematic information of single outgoing lepton and (ii) Calorimetric Method: Uses the
information of all outgoing particles for the construction of neutrino energy.\\
With the upcoming neutrino experiments the statistical error is getting reduced and that draws our
attention towards the systematic errors erupting in neutrino oscillation experiments. The main error
in the neutrino energy reconstruction arises due to nuclear effects. The
neutrino energy is reconstructed from the final state measurements and further the final state
measurements strongly depends on nuclear effects and nuclear properties. The neutrino nucleon
interaction products can be significantly modified by the presence of other nucleons inside the
nucleus. The initial interaction products can interact with other nucleons on their way and change
the particles observed in the detector from those particles which were initially produced. 
This is known as final state interaction and these interactions gives rise to fake events
in a particular neutrino nucleon interaction channel. The current uncertainties on nuclear effects
including final state interaction is quite high and needs to be addressed in upcoming neutrino
oscillation experiments like DUNE , Super-kamiokande and INO.\\
Pure events in a particular interaction channel will be those events in which the particles detected by the detector
are same as they were produced during the initial interaction. If all the events
detected by the detector are pure events then the systematic error in energy reconstruction will be
reduced to a desirable extent. But in reality a particular neutrino nucleon interaction channel events
as detected by the detector are a mixture of pure and fake events.\\
Many experiments are trying to reduce mis-reconstruction of neutrino energy by using near
detector and implementing some correlation with far detector i.e. T2K Collaboration \cite{8}.\\
Due to the mis-reconstruction of the neutrino energy the quantitative estimation of neutrino
oscillation parameters in neutrino oscillation experiment also carries some uncertainties. Number
of studies have been conducted to study the impact of nuclear effects on determination of
oscillation parameters. The impact of different models for neutrino nucleon cross section
determination on oscillation parameters is analysed in different work \cite{9,10,11}, whereas few
studies are based on quantitating the impact of final state interactions on the estimation of neutrino
oscillation parameters \cite{12,13,14}.\\
In this work we have tried to provide a quantitative estimate that final state interactions may induce
in the determination of atmospheric neutrino oscillation parameters at DUNE.

 \vspace{-3mm}
\section{DUNE flux and Neutrino Oscillation Physics:}
 The future long baseline neutrino experiment, DUNE \cite{17} (Deep Underground
Neutrino Facility) is a broadband neutrino beam experiment. The DUNE
LBNF (Long Baseline neutrino Facility) at Fermi Lab will sent intense
beam of neutrinos to the near and far detector sites. The neutrinos
generated at the LBNF will travel 1300 km to reach the high performance
DUNE far detector at Sanford Lab. The DUNE Collaboration with the use
of Long Baseline Neutrino Facility and the high performance detectors of
DUNE, aims to resolve the puzzles of neutrino with broad sensitivity to
neutrino oscillation parameters. For the generation of intense neutrino
beam for DUNE, the LBNF will extract a proton beam from the Fermilab
Main Injector (MI) and then protons will be smashed to the target where
the collisions will generate a beam of charged particles which will decay
into neutrinos to generate the neutrino beam aimed at the near and far
detectors. The LBNF is designed for initial operation at
proton beam power of 1.2 MW which will be subsequently upgraded to
proton beam power of 2.4 MW \cite{15}.\\
For this work we have selected an optimized beam flux of 120 GeV \cite{16}. The massive, deep-underground far
detector will be a 40 Ktons of liquid argon detector and it will be sensitive
for muon neutrino-appearance and muon neutrino-disappearance modes, for proton decay
modes and for observation of electron neutrinos from a core-collapse
supernova. The Fig. [1] shows the energy distribution of selected neutrino
flux and corresponding neutrino oscillation probability at 1300 km for both
muon appearance and disappearance channel.\\

\begin{figure}[H]
\begin{center}
 \includegraphics[width=14cm,height=10cm,keepaspectratio]{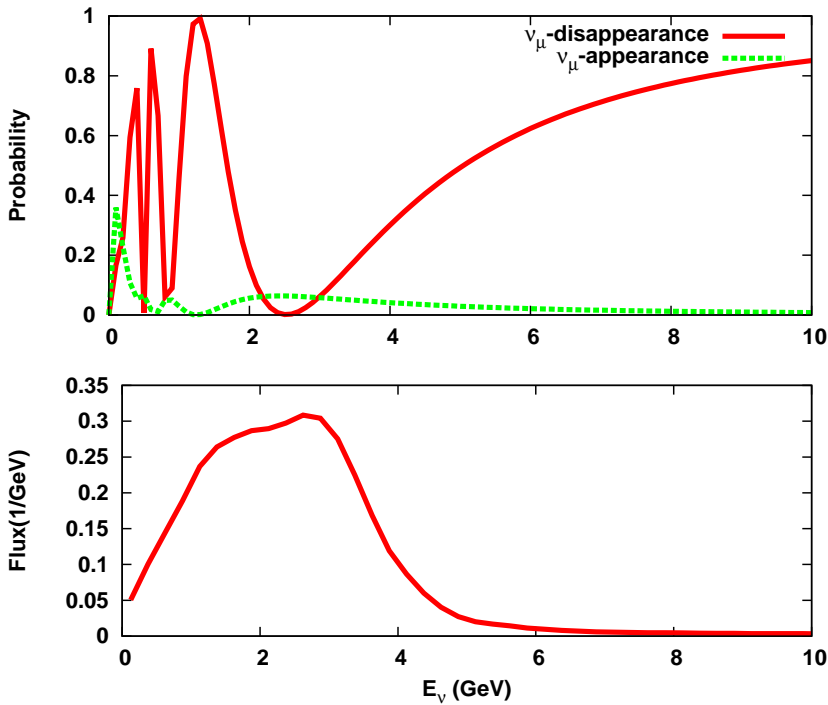}
\caption{Flux distribution for LBNE, normalized in the full energy regime (lower part) and neutrino oscillation
probability for $\nu_{\mu} \rightarrow \nu_{\mu}$ and $\nu_{\mu} \rightarrow  \nu_{e}$ channel (upper part). }
\label{1}
\end{center}
\end{figure}

The DUNE neutrino beam peaks around 2.5 GeV and in this energy region resonance
production is main mode of neutrino nucleon interaction which mainly
results in the production of pions. The dominant contribution comes
from single pion production mediated by the $\Delta (1232)$ resonance. The
charged current resonance interaction process can be illustrated as:\\
\begin{center}
                $\nu_{\mu} + p \rightarrow \mu^{-} + \Delta^{++} \rightarrow \mu^{-} + p + \pi^{+}$ \\
                $\nu_{\mu} + n \rightarrow \mu^{-} + \Delta^{+} \rightarrow \mu^{-} + n + \pi^{+}$  \\
                $\nu_{\mu} + n \rightarrow \mu^{-} + \Delta^{+} \rightarrow \mu^{-} + p + \pi^{0}$  \\
\end{center}
 The pions emitted in the resonance process can be absorbed or they can
interact with other nucleons of the nucleus resulting in an event which
gives an impression of different interaction process (fake events). In this work we have
estimated the percentage of fake events present in different neutrino nucleons interaction channel and the impact 
arising from these fake events on the estimation of
neutrino oscillation parameters.

\section{Nuclear Effects and Final State Interaction:}
Along with the large uncertainties on total neutrino cross sections the energy dependance
and energy distribution of secondary particles also adds up to the systematic errors.
One of the probable reason for this can be presence of bounded nucleon in the target
nucleus. The intense neutrino beams in upcoming experiments will greatly increase the
statistics and reduce the statistical uncertainty. Now this demands a careful handling to systematic uncertainties
in these experiments. One of the main
source of systematic uncertainty is the nuclear effects present inside the nucleus.\\
In modern day neutrino experiments complex nuclei are used as neutrino targets. These complex nuclear targets
result in non negligible or sizeable nuclear effects. Nuclear effects are taken into
consideration by Impulse Approximation \cite{19} or Fermi Gas Model \cite{20} and used by most of the neutrino experiments 
when
simulating neutrino scattering events. Many other independent particle approaches are
also developed to include spectral function \cite{21,22,23,24,25} Random Phase Approximation \cite{18,26,27,28,29} , 
a plane-wave impulse
approximation (PWIA) where the nucleon-nucleon correlations were included using description of nuclear dynamics, based on
nuclear many-body theory \cite{30} and many others. After looking at the experimental results of different neutrino experiments
performing in different energy region i.e. MINERvA, MiniBooNE, T2K,NOvA we currently
believe that nuclear effects needs to be modelled carefully.\\
The lower energy neutrino experiments like MiniBooNE and MicroBooNE are sensitive to
only two type of neutrino interactions i.e. QE and resonance. In these experiments the pion
production constitutes the one third of the total neutrino nucleon interaction cross section.
As we move towards the higher energy neutrino beam experiments NOvA, MINERvA and
ultimately, DUNE the contribution from the pion production to the total cross section
increases to the two third \cite{31}. It is thus important to have the
pion production channels under good, quantitative control.\\
Final state interaction is one of the most relevant nuclear effects that gives rise to fake
events i.e. the final particles that emerge out from the nucleus at the end of the neutrino
nucleon interaction and if detected by the detector are very different from the particles
which were formed by the neutrino nucleon interaction at initial stage. In this way the
nuclear effects can mask the true type of interaction to a different interaction type (fake),
depending on the emerging particles from the nucleus. These fake events are also known
as crossed channel events because these events cross the initial interaction channel and
give an impression of some other channel as their interaction channel, due to the presence
of nuclear effects.\\
Looking at the importance of the good knowledge on pion production channels in high
intensity neutrino beam experiments. In this work we have estimated the contribution of
nuclear effects (final state interaction) in resonance channel on the estimation of the
atmospheric neutrino oscillation parameters at DUNE.\\

\section{Simulation and Results:} 
In this work we have used GiBUU (Giessen Boltzmann-Uehling-Uhlenbeck) \cite{33} transport model to generate the events, 
cross section and migration matrices. In GiBUU, the final state interaction is modeled by solving Boltzmann-Uehling-Uhlenbeck 
(BUU) equations. The BUU equation, is the time evolution of the Wigner transform of the real-time Green's function, which is a
generalized form of classical phase-space density. In the quasi-particle limit the transport equation for Green's function, 
known as Kadanoff-Baym equation \cite{42}, contains a back-flow term that is non solvable. Botermans-Malfliet \cite{43} approximations 
and limits are used to achieve the theoretical basis for off-shell  transport model and workable BUU transport equation. 
These equations are used in Monte-Carlo simulations. For each particle species, GiBUU generate an additional differential
equation. All these equations are coupled through the relativistic mean-field (RMF) and collision terms. GiBUU includes all 
the processes e.g. true quasi-elastic, all type of resonances: delta and higher, 2p2h excitations, background pions and deep 
inelastic scattering (more detailed treatment can be found in Ref.\cite{44}).\\
In the begining we have estimated the percentage of fake events in CCQE and CCRes interaction channels. To get a quantitative
handled over this we have generated CCQE and CCRes events in the absence and presence of FSI. This is achieved
by counting the number of initial interaction
events i.e at time t=0 in a specific interaction channel “i”, N$_{p}$(i), and the number of
those events that remains in the channel after time t = 120
(intranuclear cascade is completed), N$_{f}$(i). The number of crossed channels events or fake events
are estimated by subtracting the above two values.\\

\begin{table}[H]
\caption{Threshold kinetic enegry cuts for particles }
\centering
\begin{tabular}{|c| c| c| c| c| c| c|}
\hline
Particle Type & p & n & $\pi^{+}$ & $\pi^{-}$ & $\pi^{0}$ & $\mu$\\[1.0ex]
\hline
Threshold kinetic enegry(GeV) & 0.05 & 0.05 & 0.1 & 0.1 & 0.1 & 0.03\\[1ex]
\hline
\end{tabular}
\label{1}
\end{table}

Roughly 2 lacks events are generated using DUNE flux for muon disappearance
channel. Fig. [2] and Fig. [3] represents the distribution of true and fake events as a
function of neutrino energy when no detector cuts are imposed on the out coming particles.
We can see a significant contribution arising
from the fake events in both type of interactions. The larger is the contribution from
fake events the larger are the nuclear effects on that particular interaction channel.
The percentage of fake events produced is also
calculated for both the interaction channels i.e. CCQE and CCRes and is illustrated
in Table [2].\\

\begin{table}[H]
\caption{Percentage of fake events without detector cuts}
\centering
\begin{tabular}{|c| c|}
\hline
Interaction Type & Percentage of fake events\\[1.0ex]
\hline
QE & 43.0\\
Res & 67.2 \\[1ex]
\hline
\end{tabular}
\label{2}
\end{table}

\begin{figure}[H]
\centering
\captionsetup{width=0.4\linewidth}
\begin{minipage}{.5\textwidth}
  \centering
  \includegraphics[width=9cm,height=9cm,keepaspectratio]{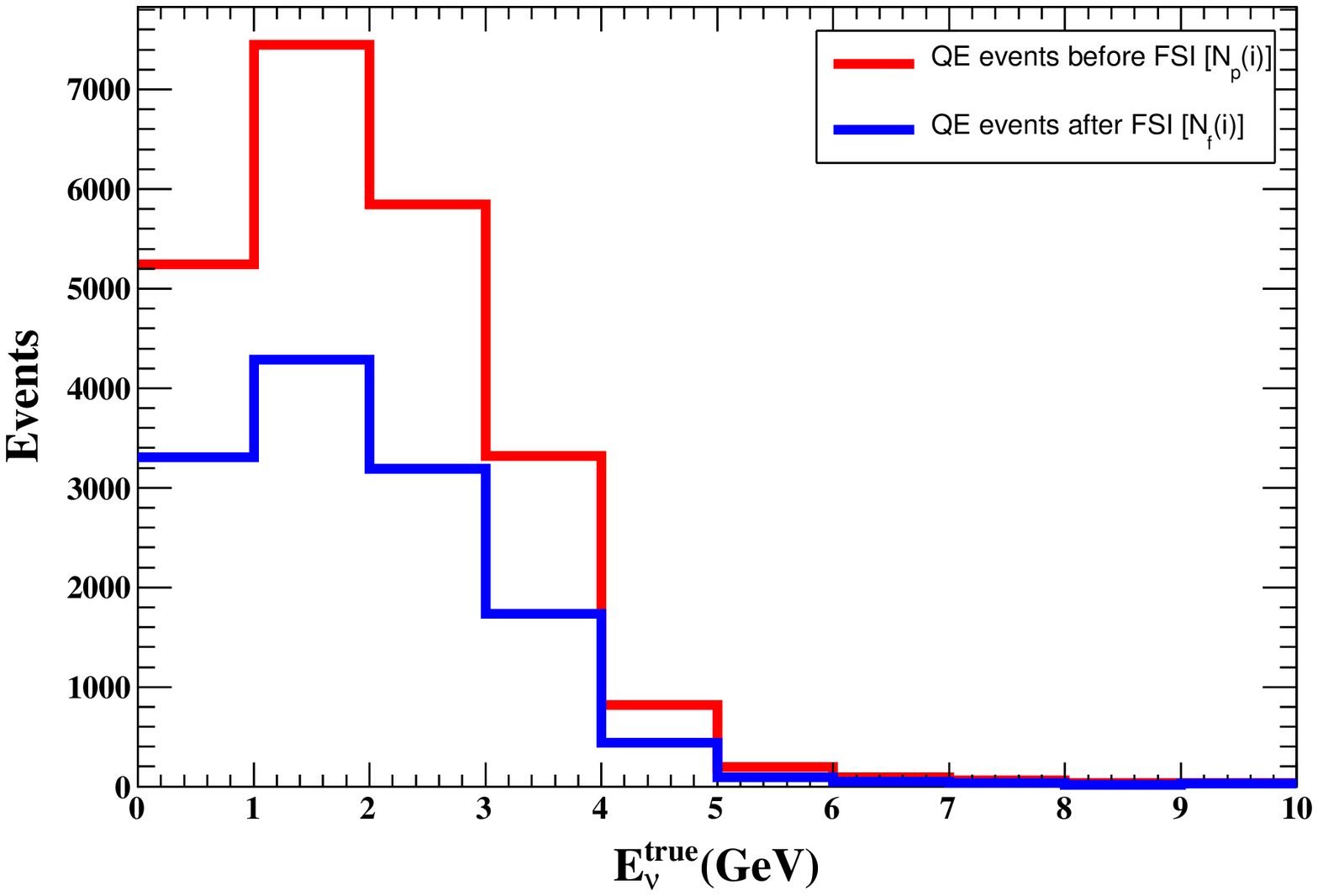}
  \caption{Distribution of QE events before (red solid line) and after (blue solid line) FSI as a function of true neutrino
energy in the absence of detector cuts.}
  \label{2}
\end{minipage}%
\begin{minipage}{.5\textwidth}
  \centering
  \includegraphics[width=9cm,height=9cm,keepaspectratio]{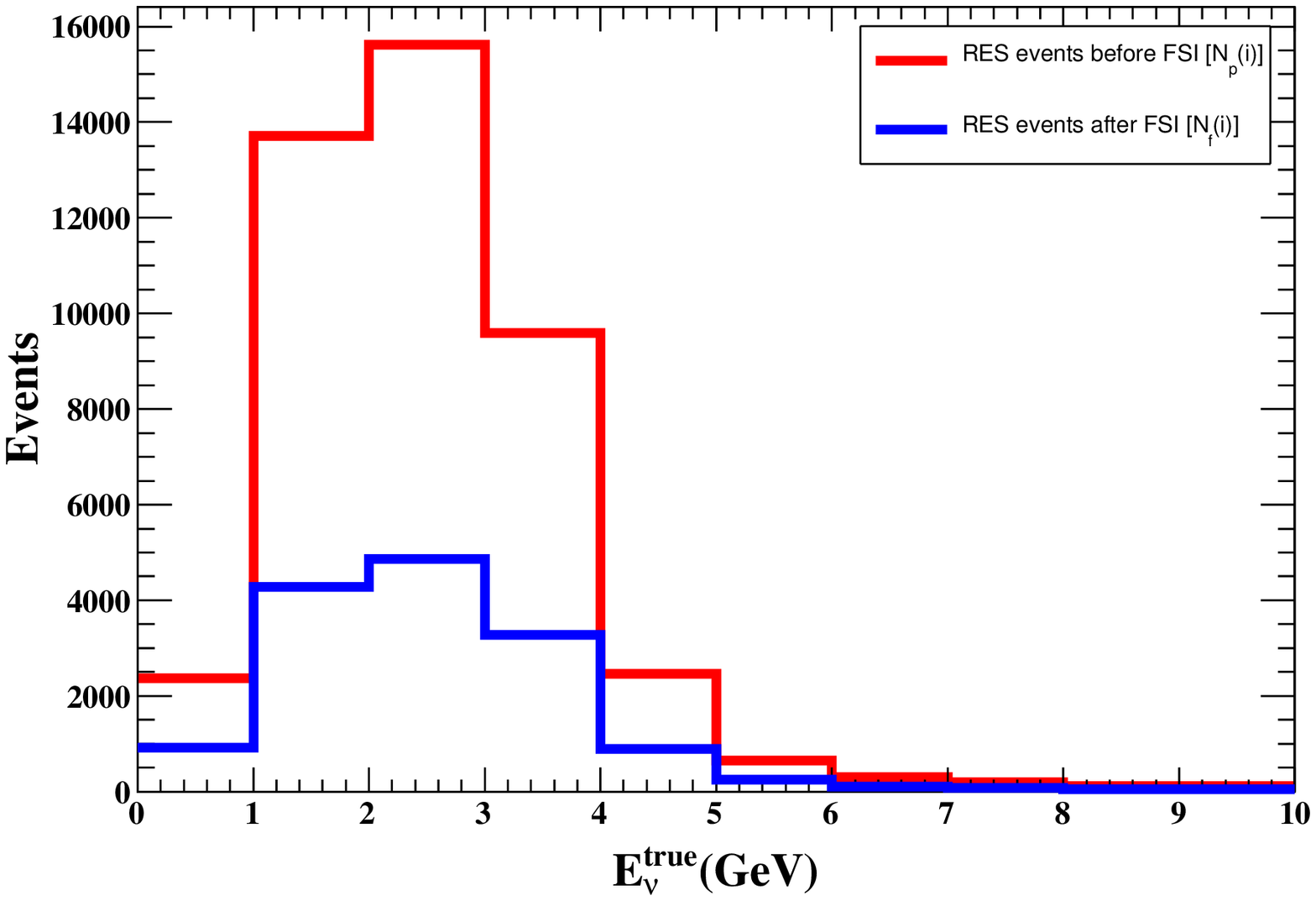}
  \caption{Distribution of Res events before (red solid line) and after (blue solid line) FSI as a function of true neutrino
energy in the absence of detector cuts.}
  \label{3}
\end{minipage}

\end{figure}

The detector cuts \cite{40} mentioned in Table [1] are imposed on the
outgoing particles to get a realistic picture of fake and true event in a
particular interaction channel. Fig. [4] and Fig. [5] represents the distribution of fake
and true events with respect to the neutrino energy after detector cuts are applied on the
data sample. The percentage contribution of the fake events are shown in Table [3].
This fact must be revised when we will include in our simulation more channels
coming from resonances higher than $\Delta$(1232).\\
 
\begin{table}[H]
\caption{Percentage of fake events with detector cuts}
\centering
\begin{tabular}{|c| c|}
\hline
Interaction Type & Percentage of fake events\\[1.0ex]
\hline
QE & 51.05\\
Res & 72.4 \\[1ex]
\hline
\end{tabular}
\label{3}
\end{table} 
  
\begin{figure}[H]
\centering
\captionsetup{width=0.4\linewidth}
\begin{minipage}{.5\textwidth}
  \centering
  \includegraphics[width=9cm,height=9cm,keepaspectratio]{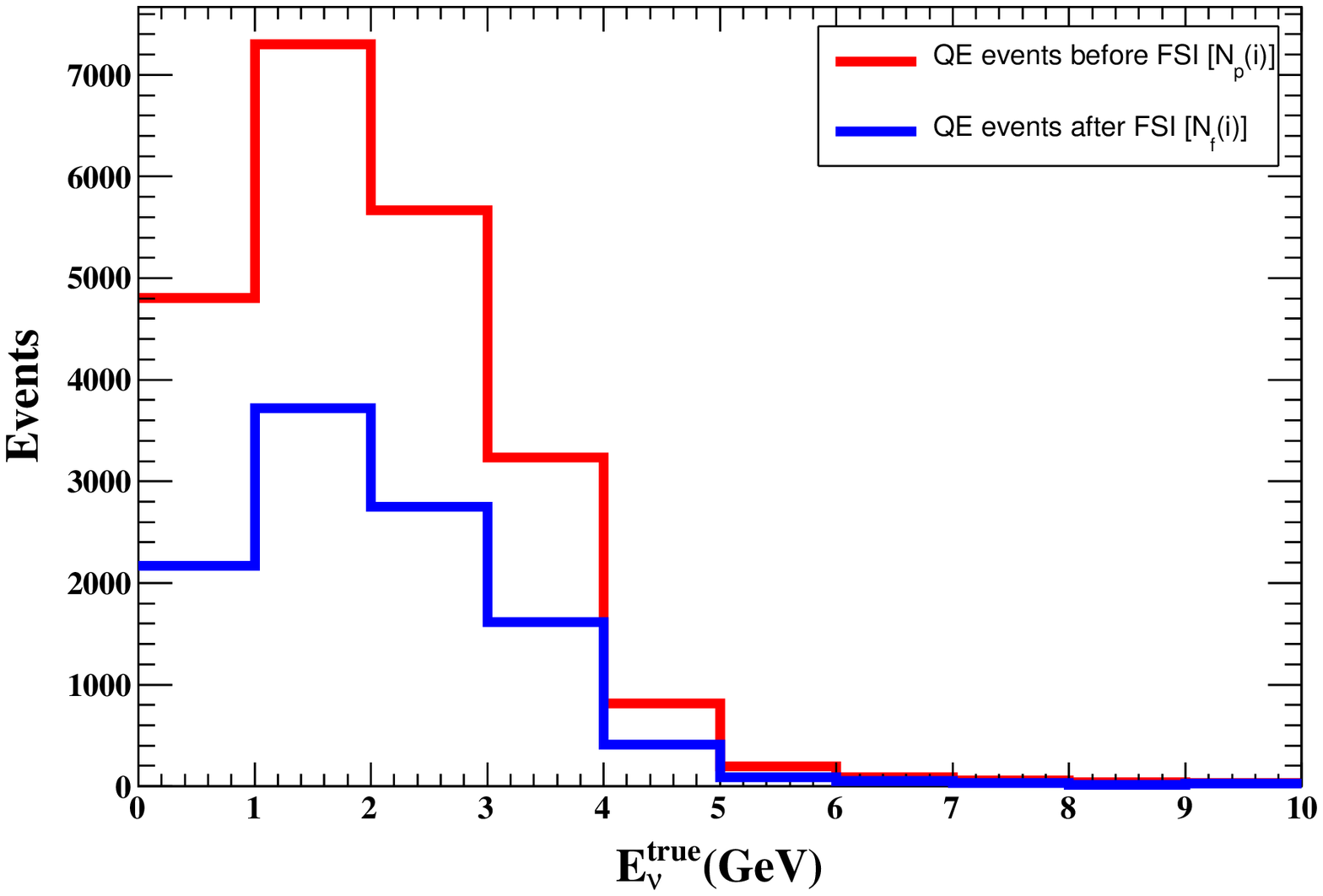}
  \caption{Distribution of QE events before (red solid line) and after (blue solid line) FSI as a function of true neutrino
energy in the presence of detector cuts.}
  \label{4}
\end{minipage}%
\begin{minipage}{.5\textwidth}
  \centering
  \includegraphics[width=9cm,height=9cm,keepaspectratio]{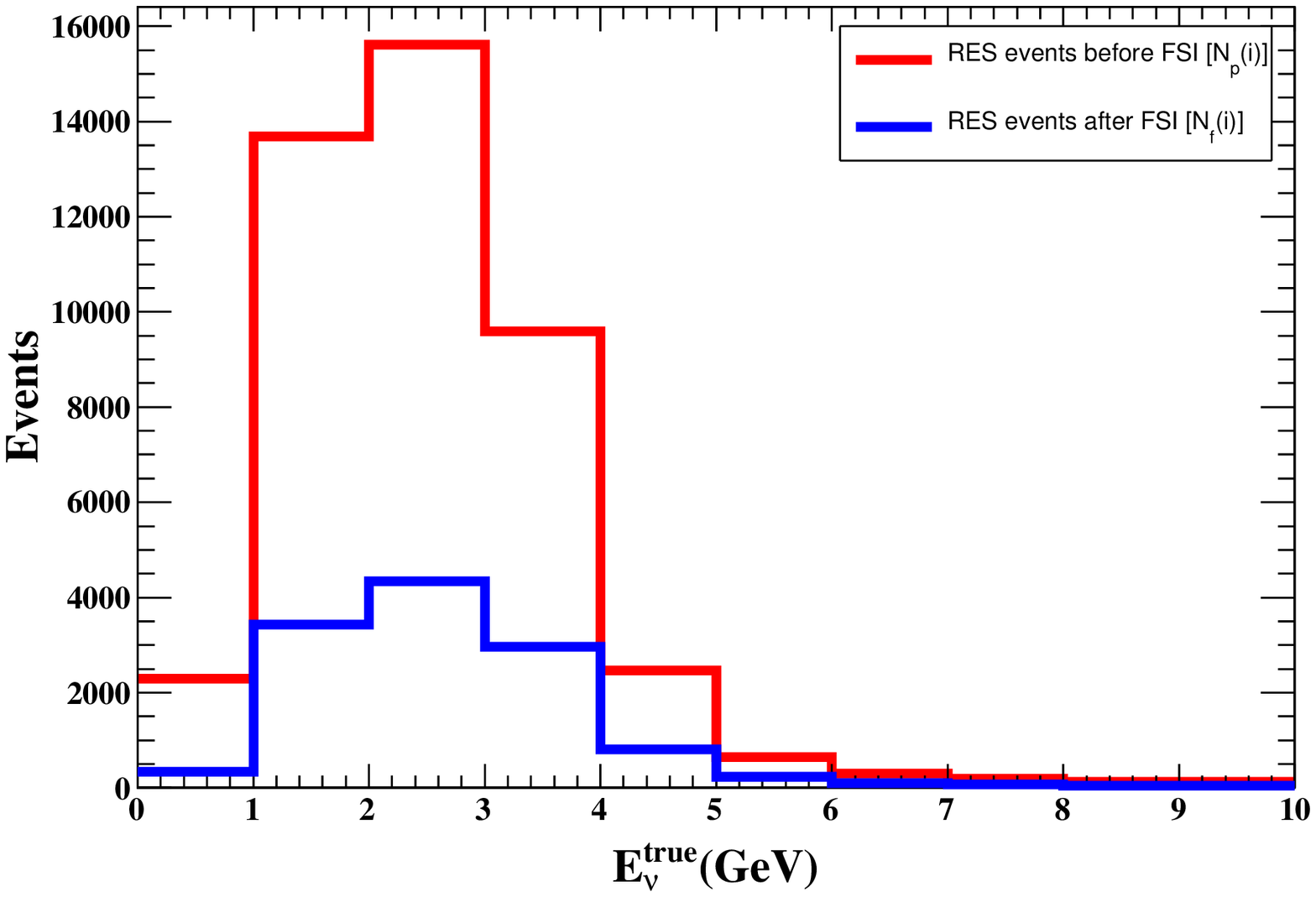}
  \caption{Distribution of Res events before (red solid line) and after (blue solid line) FSI as a function of true neutrino
energy in the presence of detector cuts.}
  \label{5}
\end{minipage}

\end{figure}

From the previous plots it is visible that the contribution arising from the fake events to the total events gathered in a 
charged current neutrino nucleon interaction is more than 50\% in both cases. With such a markable contributions arising from 
the nuclear effects a reanalysis of neutrino oscillation parameters in presence of nuclear effects becomes crucial. As we know 
that the neutrino oscillation physics depends on the values of neutrino oscillation parameters and the value of these 
oscillation parameters depends on the precise knowledge of neutrino energy. Due to the presence of nuclear effects a difference 
between true and reconstructed energy of neutrino becomes sizable and this needs to be pined down with the knowledge of nuclear effects.
To include the nuclear effects in our work a migration matrix between true and reconstructed energy with a bin size of 0.5 GeV 
is generated using GiBUU. To quantitate the effects of final state interaction on neutrino oscillation physics the BUU 
migration matrix and cross sections estimated using GiBUU are included in correct format to the latest version of 
GLoBES(Global Long Baseline Experiment Simulator)\cite{32, 34, 35, 36, 37, 38}. In this work the effects of final state interaction 
at DUNE oscillation physics has been estimated and for this we use the DUNE neutrino flux and liquid argon far detector of
fiducial mass 35 Kt, placed at a baseline of 1300 Km. This work is performed for muon disappearance channel and running 
time considered here is 5 years in neutrino mode and 5 years in antineutrino mode. The true values of
oscillation  parameters proposed  in  this work  are $\theta_{23}$ =
 45.0$^{0}$, $\theta_{12}$ = 34.5$^{0}$, $\theta_{13}$ = 8.44$^{0}$, $\Delta m^{2}_{31}$ = 2.55 $\times$ 10$^{-3}$ eV$^{2}$, 
 $\Delta m^{2}_{21}$ = 7.56 $\times$ 10$^{-5}$ eV$^{2}$ and $\delta_{CP}$ = 0$^{0}$ \cite{39}.
The signal efficiency is 85\% for disappearance channel , normalization error of signal and background are 5\% and 10\% 
respectively and energy calibration error of signal and background are 2\%. The migration matrix for pure QE and Res events 
and QE-like and Res-like events is obtained useing GiBUU. Resonance like events are those events which at  the initial point of 
neutrino nucleon interaction were not resonance events instead they were something different i.e. QE, DIS or something else, 
but while leaving the nucleus they appear like if they were produced due to resonance interaction.
Let us consider now the case of a charge current neutrino interaction that is not Resonance. Usually, these interactions are discarded 
from the event sample if another charged particle (for example, a pion) is not observed in the final stale.
However, there is a certain probability that the produced hadron while passing through the nucleus interacts with other 
nucleon and produces a pion. In this case, pion will be observed in the final slate and consequently this events will be added 
to the resonance sample. The difference in true and reconstructed neutrino energy arises due to these events.\\
Estimation of the final stale interaction on neutrino oscillation parameters two extreme cases are considered
(i)when nuclear effects are completely disregarded and
(ii)when  nuclear  effects  are  completely known.
This  condition  is  achieved  by  plugging  a
parameter $\alpha$ to true and like events which eventually adds up to give the total number of events in a particular 
interaction channel.
The parameter $\alpha$ can take any vaue between 0 and 1. In reality we lie somewhere in between these two 
situations of complete ignorance and complete knowledge of nuclear effects. Total number of events can be represented as:\\
\begin{equation}
 \label{1}
  N^{test}_{i}(\alpha) = \alpha \times N^{QE}_{i} + (1 - \alpha) \times N^{QE-like}_{i}
 \end{equation}
\begin{equation}
\label{2}
  N^{test}_{i}(\alpha) = \alpha \times N^{RES}_{i} + (1 - \alpha) \times N^{Res-like}_{i}
 \end{equation}
 1. When $\alpha$ = 1 (nuclear effects are completely disregarded).\\
 2. When $\alpha$ = 0 (nuclear effect are perfectely known).\\  
      
The inclusion of the parameter $\alpha$ can be considered as systematic uncertainty, Similar
approach is also considered in \cite{41}.

\begin{figure}[H]
\centering
\captionsetup{width=0.4\linewidth}
\begin{minipage}{.5\textwidth}
  \centering
  \includegraphics[width=9cm,height=9cm,keepaspectratio]{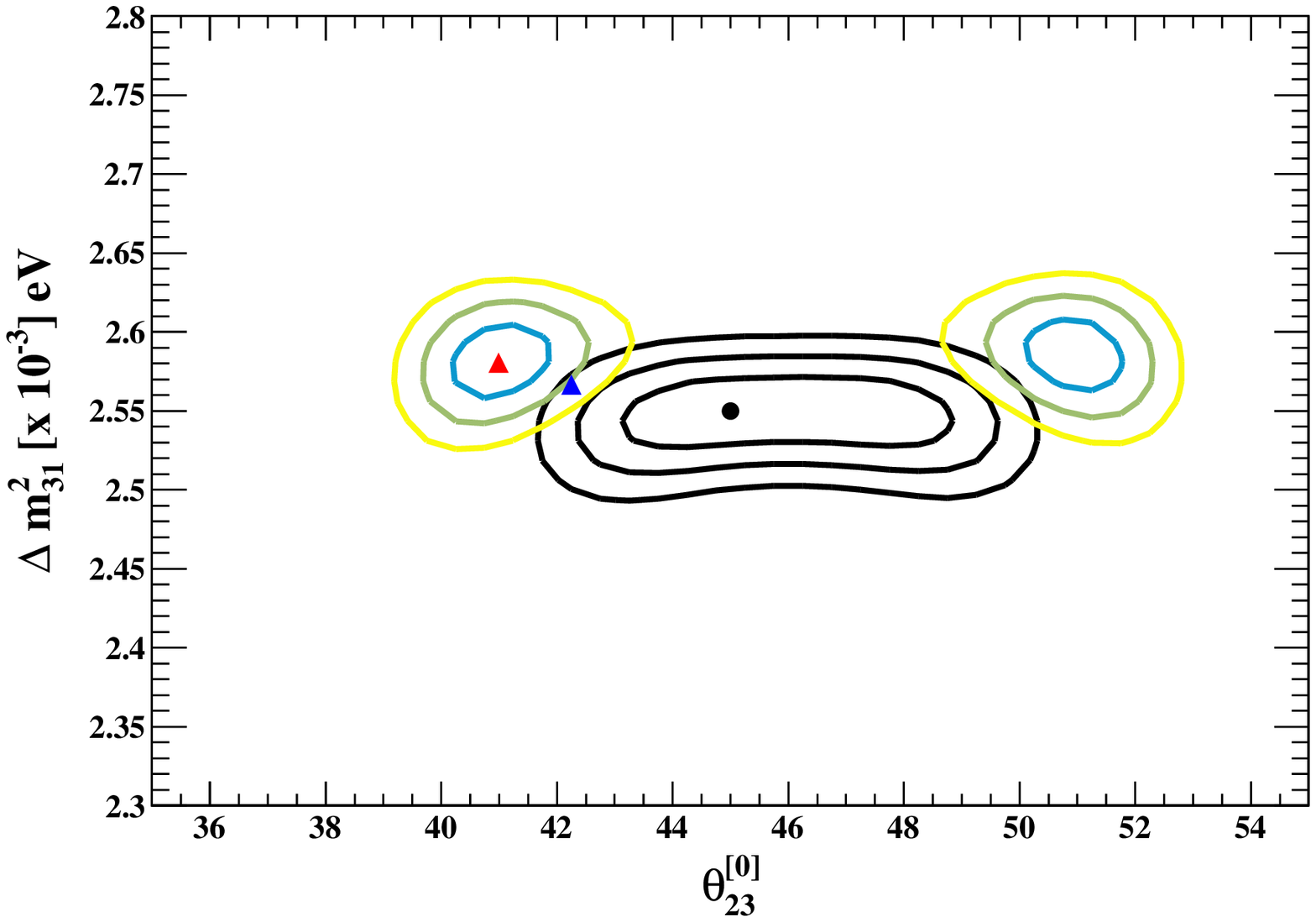}
  \caption{Confidence regions in the
($\theta _{23}, \Delta m_{31}^{2}$) plane are obtained using the migration matrices pure QE (black lines) and 
QE-like (color lines) in the absence of detector cuts. The red triangle($\alpha = 0$), blue triangle($\alpha = 0.5$) shows the best
fit point and circle($\alpha = 1$) show the true values of the oscillation parameters.}
  \label{6}
\end{minipage}%
\begin{minipage}{.5\textwidth}
  \centering
  \includegraphics[width=9cm,height=9cm,keepaspectratio]{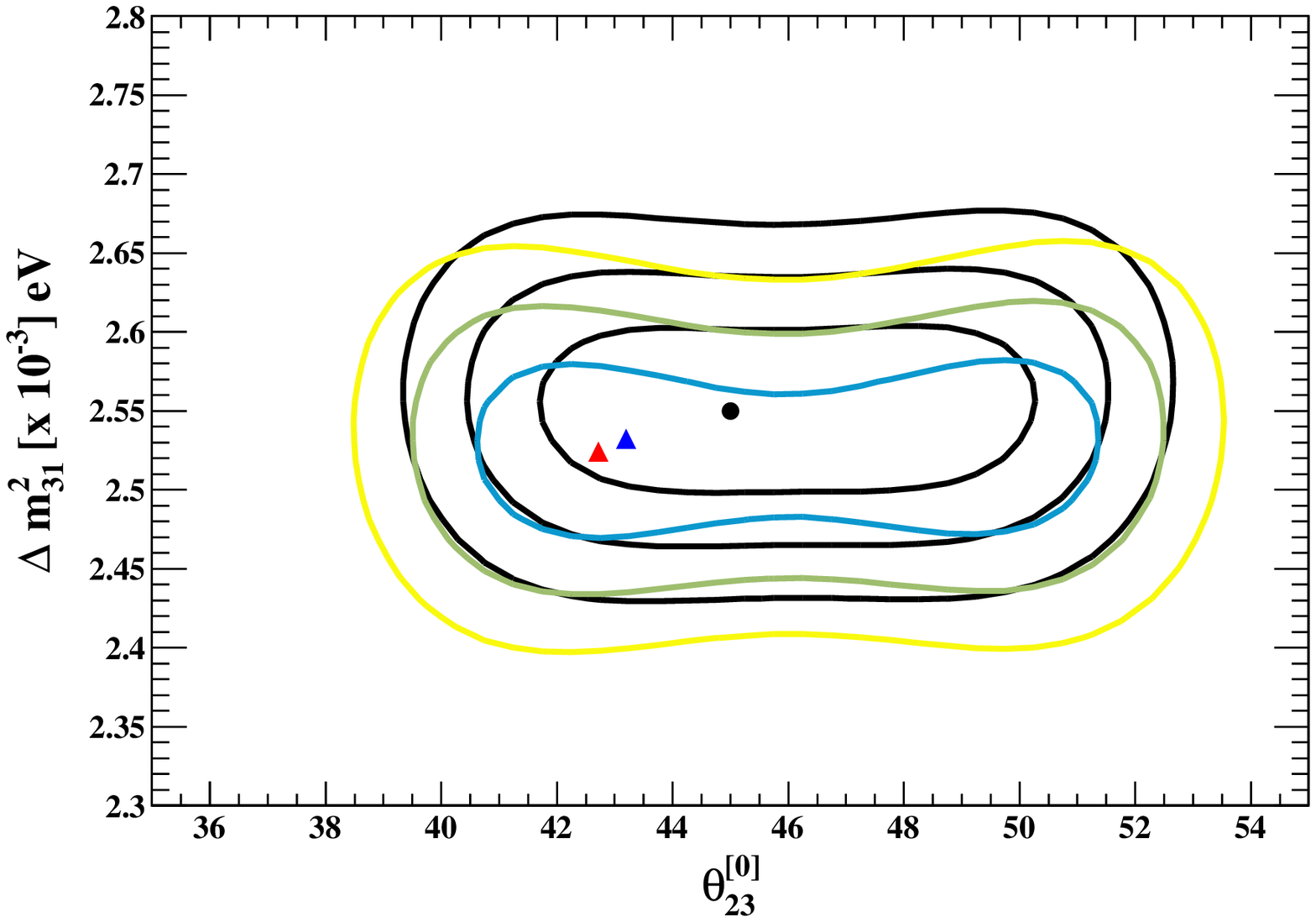}
  \caption{Confidence regions in the
($\theta _{23}, \Delta m_{31}^{2}$) plane are obtained using the migration matrices pure Res (black lines) and 
Res-like (color lines) in the absence of detector cuts. The red triangle($\alpha = 0$), blue triangle($\alpha = 0.3$) shows the best
fit point and circle($\alpha = 1$) show the true values of the oscillation parameters.}
  \label{7}
\end{minipage}

\end{figure}

A chi square plot between two oscillation parameters $\Delta m_{31}^{2}$ and $\theta _{23}$ for QE and Res
events in absence and presence of detector cuts are shown in Fig. [6], Fig. [7] and Fig. [8], Fig. [9]
respectively. The 1, 2, and 3$\sigma$ confidence regions in the ($\theta _{23}$ , $\Delta m_{31}^{2}$) plane are obtained
using the migration matrices accounting for (coloured lines) or neglecting (black lines)
the effect of final-state interactions, included in the simulated data. The coloured triangle
and the black circle show the best fit point and the true values of oscillation parameters
respectively.\\

\begin{figure}[H]
\centering
\captionsetup{width=0.4\linewidth}
\begin{minipage}{.5\textwidth}
  \centering
  \includegraphics[width=9cm,height=9cm,keepaspectratio]{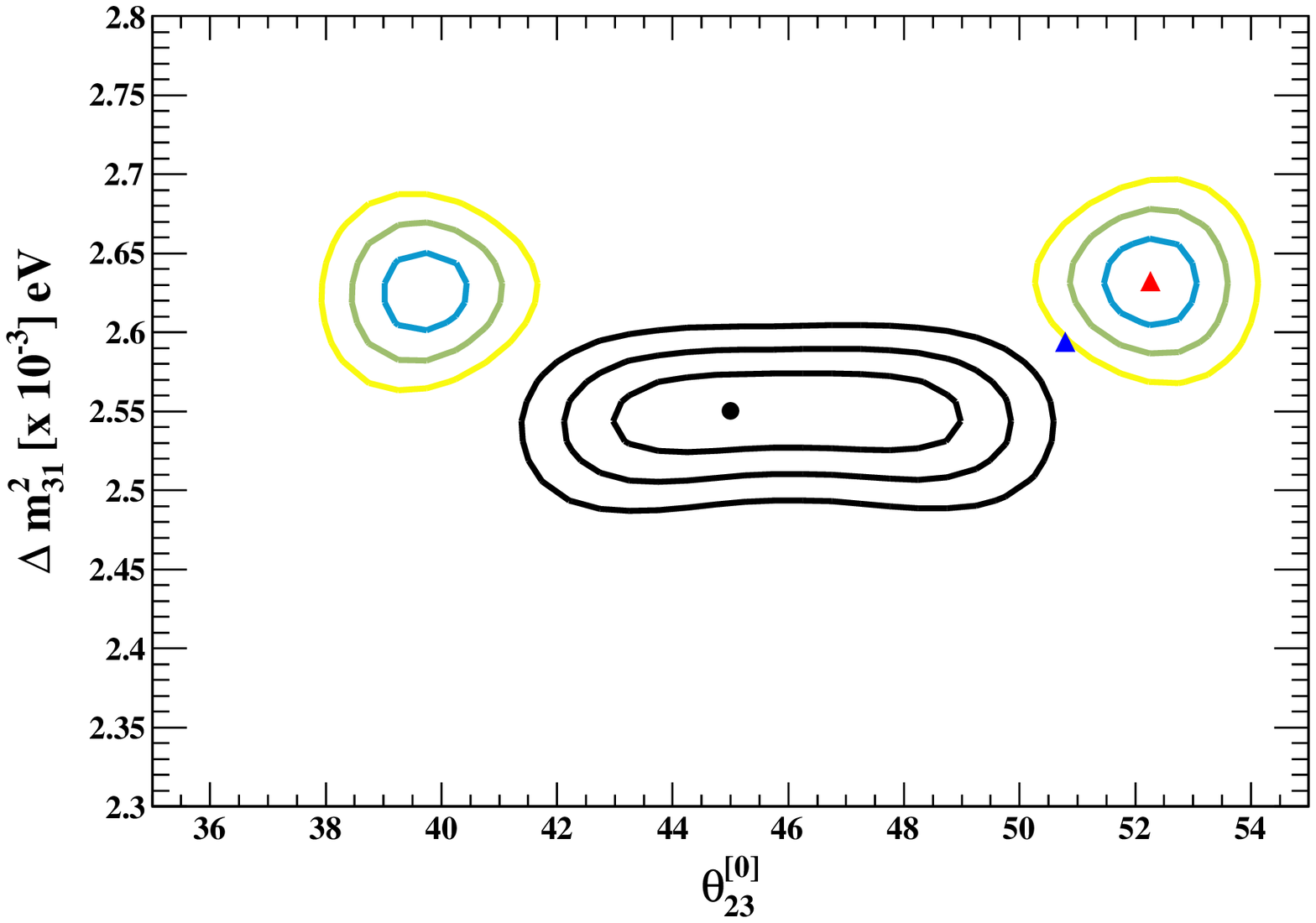}
  \caption{Confidence regions in the
($\theta _{23}, \Delta m_{31}^{2}$) plane are obtained using the migration matrices pure QE (black lines) and 
QE-like (color lines) in the presence of detector cuts. The red triangle($\alpha = 0$), blue triangle($\alpha = 0.5$) shows the best
fit point and circle($\alpha = 1$) show the true values of the oscillation parameters.}
  \label{8}
\end{minipage}%
\begin{minipage}{.5\textwidth}
  \centering
  \includegraphics[width=9cm,height=9cm,keepaspectratio]{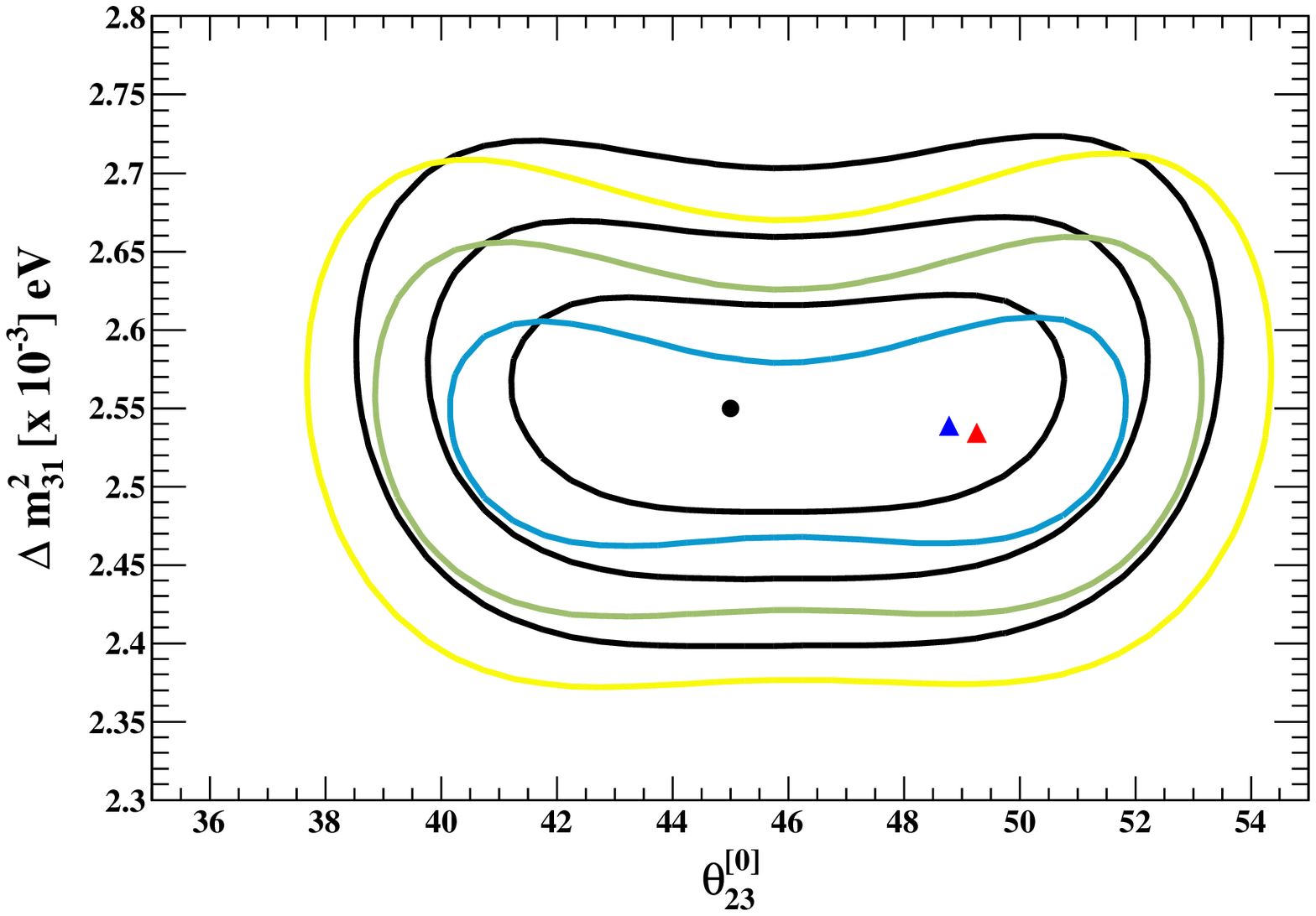}
  \caption{Confidence regions in the
($\theta _{23}, \Delta m_{31}^{2}$) plane are obtained using the migration matrices pure Res (black lines) and 
Res-like (color lines) in the presence of detector cuts. The red triangle($\alpha = 0$), blue triangle($\alpha = 0.3$) shows the best
fit point and circle($\alpha = 1$) show the true values of the oscillation parameters.}
  \label{9}
\end{minipage}

\end{figure}
      
As estimated in Table [1] and Table [2] roughly 50\% of the QE events remains QE after final
state interaction. A best fit point for $\alpha = 0.5$ is obtained using Equ. [1] is also shown
in the Fig. [6] and Fig. [8] which refers to 50\% QE and 50\% QE-like events. Similarly the
mentioned table shows 30\% pure Res events after FSI following that a best fit point for
$\alpha = 0.3$ is obtained using Equ. [2] and is shown in Fig. [7] and Fig. [9].

\section{Conclusion:}
In the present work we report an extensive analysis of nuclear effects in neutrino-nucleus
interaction at DUNE. For this purpose, GiBBU is used to take into account the nuclear
effects. CCQE and CCRes interaction channels are analyzed in this study. Effect of FSI
on CCQE interaction channel is also studied in \cite{41}. In neutrino oscillation physics at
DUNE resonance interactions plays vital role due to the shape of DUNE flux and in this
work we have reported the effects of FSI on CCRes interaction channel too.\\
The calculations are performed with DUNE flux for liquid Argon detector. We calculated
the cross section of detector nuclei in the whole energy range of DUNE flux using CCQE
and CCRes channels. For the given target nuclei, we perform an exploratory study of
the fake events generated in several reactions. The percentage of fake events increases in
resonance events and they increase further when we use the real detector (imposing detector cuts). In a future
work, we will improve the work by adding higher resonance and DIS studies to it.\\
The position of the best fit corresponding to different values of $\alpha \sim$ 0, 0.5, 1 for CCQE are
shown in the Fig. [6] and Fig. [8] whereas in Fig.[7] and Fig. [9] the best fit corresponding to $\alpha \sim$ 0,
0.3, 1 for CCRes are illustrated. As it can be seen in the mentioned figures, the deviation
of the best fit from the true input value gets progressively increased with the increase in the
value of $\alpha$ or when we move from nuclear effects completely disregarded to nuclear
effects completely known. However, according to Fig. [6] for relatively large value of
$\alpha \sim$0.5 the minimum value of $\chi ^{2}$ for CCQE would correspond to a $3\sigma$ bias in the
determination of the mixing angle though according to Fig. [8] it goes beyond the $3\sigma$ limit
when detector cuts are taken into consideration. In Fig. [7] and Fig. [9] the best fit value for
$\alpha \sim$ 0.3 for CCRes (with and without detector cuts) corresponds to $1\sigma$ limit in the
determination of mixing angles. Therefore, it stands to reason that a successful
experiment requires an accurate nuclear model, where the accuracy of the model has been
independently verified.\\
Our results indicate that, for an experiment observing most of Res-like events, a $1\sigma$ bias in
the determination of $\theta _{23}$ and for a experiment observing most of QE-like events roughly a
$3\sigma$ bias in the determination of $\theta _{23}$ could result from errors on the nuclear model. In an
outlook of the study we can conclude that the best strategy for third generation neutrino-
oscillation experiments seems to minimize detection thresholds of the employed
detectors and to perform an extensive authentication of the accuracy of nuclear models
employed in data analysis.

\begin{acknowledgement}
we are deeply indebted to Prof. Raj Gandhi for his invaluable help at all stages of this
project and for his comments on the manuscript.

\end{acknowledgement}\vspace{-10mm}
%


\end{document}